\documentclass[aps,prc,twocolumn,superscriptaddress]{revtex4-2}
\usepackage{xcolor}
\usepackage{graphicx}
\usepackage{amsmath}
\usepackage{overpic}
\usepackage{booktabs} 

\begin{document}

\title{Deciphering the universal scaling of particle transverse momentum spectra in heavy-ion collisions}

\author{Xi-Yao Guo}
\affiliation{School of Physics and Information Technology, Shaanxi Normal University, Xi'an 710119, China}
\author{Hua Zheng}
\email[Corresponding author: ]{zhengh@snnu.edu.cn}
\author{Wenchao Zhang}
\affiliation{School of Physics and Information Technology, Shaanxi Normal University, Xi'an 710119, China}
\author{Li-Lin Zhu}
\affiliation{College of Physics, Sichuan University, Chengdu 610064, China}
\author{Xing-Quan Liu}
\affiliation{Key Laboratory of Radiation Physics and Technology of the Ministry of Education, Institute of Nuclear Science and Technology, Sichuan University, Chengdu 610064, China}
\author{Zhi-Guang Tan}
\affiliation{School of Electronic Information and Electrical Engineering, Changsha University, Changsha 410003, China}
\author{Dai-Mei Zhou}
\affiliation{Key Laboratory of Quark Lepton Physics (MOE) and Institute of Particle Physics, Central China Normal University, Wuhan 430079, China}
\author{Ben-Hao Sa}
\affiliation{Key Laboratory of Quark Lepton Physics (MOE) and Institute of Particle Physics, Central China Normal University, Wuhan 430079, China}
\affiliation{China Institute of Atomic Energy, P. O. Box 275 (10), Beijing 102413, China}

\date{\today}

\begin{abstract}
We systematically investigate the scaling properties of the transverse momentum spectra for pions, kaons, and protons in Au+Au collisions at $\sqrt{s_{NN}}$ = 7.7, 11.5, 14.5, 19.6, 27, 39, 62.4, and 200 GeV, as well as in U+U collisions at $\sqrt{s_{NN}}$ = 193 GeV, across different centrality classes, using experimental data from the collaborations at the Relativistic Heavy Ion Collider (RHIC). Universal scaling emerges when the particle transverse momentum spectra are scaled by global physical quantities, i.e., the average total particle multiplicity and mean transverse momentum, confirming recent scaling findings from the data at the Large Hadron Collider (LHC) by the ExTrEMe collaboration. The scaling behavior breaks down in the high $p_{T}$ region and in peripheral collisions. We provide a natural explanation for these observations by invoking the Cooper-Frye formula, which is used for hadronization in hydrodynamics. Furthermore, we demonstrate the equivalence between the scaling found by the ExTrEMe collaboration and the Hwa-Yang scaling which was proposed two decades ago.
\end{abstract}

\maketitle

\section{Introduction}
The primary goal of high energy collision experiments in terrestrial laboratories is to explore the fundamental properties of nuclear matter under extreme conditions, i.e., high temperature and density~\cite{Hwa:2004yg,Hwa:2010npa,Csernai1994IntroductionTR,Wong1994,Chen:2024aom,He:2026fsq,Zhang2026}. Since the collision processes cannot be measured directly, experimental observables measured from the final-state particles act as essential probes to decipher their properties and dynamical evolution~\cite{Mohapatra:2016kii,ALICE:2015bpk,STAR:2006kxj,Tao:2020uzw,Tao:2023kcu,Zhao:2018lyf,Tan:2024lrp}. Among these observables, the transverse momentum $p_{\rm T}$ spectra of identified particles are useful for investigating particle collective dynamics and hadronization mechanisms in high energy heavy-ion collisions~\cite{Jena:2020wno,Fries:2003vb,STAR:2003jis,Gupta:2020naz,Tao:2022tcw,Xie:2026klq}. Measurements at the RHIC and LHC have revealed a range of striking phenomena from particle transverse momentum spectra, such as the anomalous enhancement of baryon production at intermediate transverse momentum region~\cite{Lamont:2007ce,PHENIX:2013kod,ALICE:2014juv}, strong high-$p_{\rm T}$ suppression of hadrons~\cite{PHENIX:2001hpc,STAR:2003pjh}, and the number-of-constituent-quark (NCQ) scaling of elliptic flow of identified particles~\cite{Csanad:2005gv,Wang:2022det,STAR:2015gge,STAR:2004jwm,STAR:2002hbo}. These observations provide key insights into the properties and hadronization mechanisms of the strongly coupled quark-gluon plasma (QGP). 

Among various approaches to studying heavy-ion collision physics, the search for universal scaling patterns in observables plays an important role. Scaling simply implies that a universal physics exists and can deepen our understanding, such as KNO scaling~\cite{Koba:1972ng,Sarcevic:1987ry,Duan:2025ngi}, Bjorken scaling~\cite{Bjorken:1968dy,Altarelli:1977zs}, Regge Scaling~\cite{Regge:1959mz,White:1993na,Braun:2025eve}, and Feynman Scaling~\cite{Feynman:1969ej,Benecke:1969sh,Arakelyan:2012wu}. The scaling behavior of particle transverse momentum spectra has long served as a valuable probe in relativistic heavy-ion collisions \cite{Jiang:2013gxa,Hwa:2002tu,Hwa:2003zp,Hwa:2003bn}. In 2003, Hwa and Yang discovered pion transverse momentum spectra scaling, known as Hwa-Yang scaling, in heavy-ion collisions across different collision energies and centralities by introducing a scaling variable $z=p_{\rm T}/{K}$, where $K$ is a parameter that depends on the centrality and collision energy ~\cite{Hwa:2002tu,Hwa:2003zp,Hwa:2003bn}. They further showed that this scaling behavior could be used to define a KNO-like scaling function $\Psi(u)$ with $u=p_{\rm T}/\left \langle p_{\rm T} \right \rangle $, which removes the dependence on the scaling parameter and relies only on the global quantity, i.e., the mean transverse momentum $\langle p_{\rm T}\rangle$. This approach was later extended to various collision systems, particle species, and collision energies, establishing a comprehensive scaling framework, which is interpreted within the color string percolation model~\cite{Zhu:2006gv,Zheng:2008zzg,Zhu:2008zzd,Zhu:2008uu,Zhang:2007st,Zhang:2014dna,Zhang:2015wea,Yang:2017cup,Wang:2018pbh,Liu:2019fdp}. It was concluded that the scaling is universal, independent of collision energies, collision systems, and particle species. Recently, a new particle transverse momentum spectra scaling was proposed in Refs.~\cite{ExTrEMe:2024fxt, Domingues:2025tkp, Domingues:2026xvs, Domingues:2026mis}.  The variable $x_{\rm T} = p_{\rm T}/\langle p_{\rm T}\rangle$ and a scaled spectrum $U(x_{\rm T}) = (\langle p_{\rm T}\rangle/N) \cdot dN/dp_{\rm T}$, which are scaled using the global physical quantities of the average total particle multiplicity $N$ for the chosen particle species and mean transverse momentum $\langle p_{\rm T} \rangle$, are adopted. $U(x_{\rm T})$ for pions at the LHC exhibits a nearly universal, centrality-independent scaling. Such a behavior has also been observed in the hybrid hydrodynamic simulations and interpreted as a possible signature of hydrodynamic evolution~\cite{ExTrEMe:2024fxt}. Similar universal scalings are also found for kaons, and protons at the LHC~\cite{Domingues:2025tkp, Domingues:2026xvs, Domingues:2026mis}. Furthermore, a comprehensive study reveals that the hybrid hydrodynamic models have little flexibility in describing the scaled spectrum observables by varying the model parameters \cite{Domingues:2026mis}.  Therefore, the underlying physical mechanism leading to such scaling remains unclear. 

It is natural to extend this particle transverse momentum spectra scaling search to the pion, kaon, and proton spectra in heavy-ion collisions at the RHIC, as motivated by the Hwa-Yang scaling. It is expected that the scaling is universal as well. In this paper, the scaling properties of the transverse momentum spectra for pions, kaons, and protons in Au+Au collisions at $\sqrt{s_{NN}}$ = 7.7, 11.5, 14.5, 19.6, 27, 39, 62.4, and 200 GeV, as well as in U+U collisions at $\sqrt{s_{NN}}$ = 193 GeV at the RHIC are investigated. We demostrate that the observed scaling persists across this wide energy range and centrality classes at both the RHIC and LHC, and provide a natural theoretical interpretation rooted in the Cooper-Frye formula. Furthermore, we establish the relation between the scaling found by the ExTrEMe collaboration and the Hwa-Yang scaling, demonstrating their equivalence and deepening our understanding of the particle spectra scaling found in heavy-ion collisions.

The organization of this paper is as follows. In Sec.~\ref{two}, we briefly describe the procedure for searching for particle transverse momentum spectra scaling. Section~\ref{three} presents the scalings of the transverse momentum spectra for pions, kaons, and protons in Au+Au collisions at RHIC energies from 7.7 to 200 GeV and in U+U collisions at 193 GeV. A comparison with the results obtained at the LHC is also shown. In Sec.~\ref{four}, we give our explanation for the mechanism leading to the particle transverse momentum spectra scaling by using the Cooper-Frye formula, showing quantitative agreement with data. We also explicitly demonstrate the equivalence between the scaling found by the ExTrEMe collaboration and the Hwa-Yang scaling. Finally, a brief conclusion is given in Sec.~\ref{five}.

\section{Method to search for the particle transverse momentum spectra scaling}\label{two}
As done in Refs.~\cite{ExTrEMe:2024fxt,Domingues:2025tkp,Domingues:2026xvs,Domingues:2026mis}, the experimental data of particle transverse momentum spectra are converted into the transverse momentum differential distribution $dN/dp_{\rm T}$. In order to scale the particle distribution, the average total particle multiplicity $N$ and mean transverse momentum $\langle p_{\rm T}\rangle$ of a given particle species are required for a given centrality. Note that they are calculated from $dN/dp_{\rm T}$ starting from $p_{\rm T}=0$ GeV/c. 

With $N$ and $\langle p_{\rm T}\rangle$ for each centrality, the scaled transverse momentum distribution
\begin{equation}~\label{eq1}
U(x_{\rm T})=\frac{\left \langle p_{\rm T} \right \rangle}{N} \frac{dN}{dp_{\rm T}}=\frac{1}{N} \frac{dN}{d x_{\rm T}}
\end{equation}
can be derived, where $x_{\rm T}=p_{\rm T}/\left \langle p_{\rm T} \right \rangle$. The uncertainties of the scaled transverse momentum distribution $U(x_{\rm T})$ are estimated from the particle $p_{\rm T}$ spectrum errors, as well as the uncertainties of $\langle p_{\rm T}\rangle$ and $N$.

Then the scaled particle transverse momentum spectra $U(x_{\rm T})$ as function of $x_{\rm T}$ for all centralities are plotted together. If all scaled data points collapse onto a single curve, it means that the particle transverse momentum spectra scaling exists. Usually, the scaling is not perfect. The ratio 
\begin{equation}
R_i(x_{\rm T})=\frac{U_{i}(x_{\rm T})}{U_{MCC}(x_{\rm T})} \label{ratioeq}
\end{equation}
is defined to quantify any deviations, where $U_{i}(x_{\rm T})$ is the scaled spectrum for the $i$th centrality bin, and $U_{MCC}(x_{\rm T})$ corresponds to scaled spectrum for the most central centrality (MCC). Since different centrality classes have different $\langle p_{\rm T}\rangle$ values and thus different $x_{\rm T}$, we interpolate the scaled spectrum $U_{MCC}(x_{\rm T})$ onto the $x_{\rm T}$ points of each centrality bin for a point-by-point comparison. The same approach was applied in Ref.~\cite{ExTrEMe:2024fxt} to ensure a consistent quantitative evaluation of scaling violations across centralities.

\section{Results of pion, kaon and proton transverse momentum spectra scalings at RHIC}\label{three}
\begin{figure*}[!htbp]
\centering
\includegraphics[width=\textwidth]{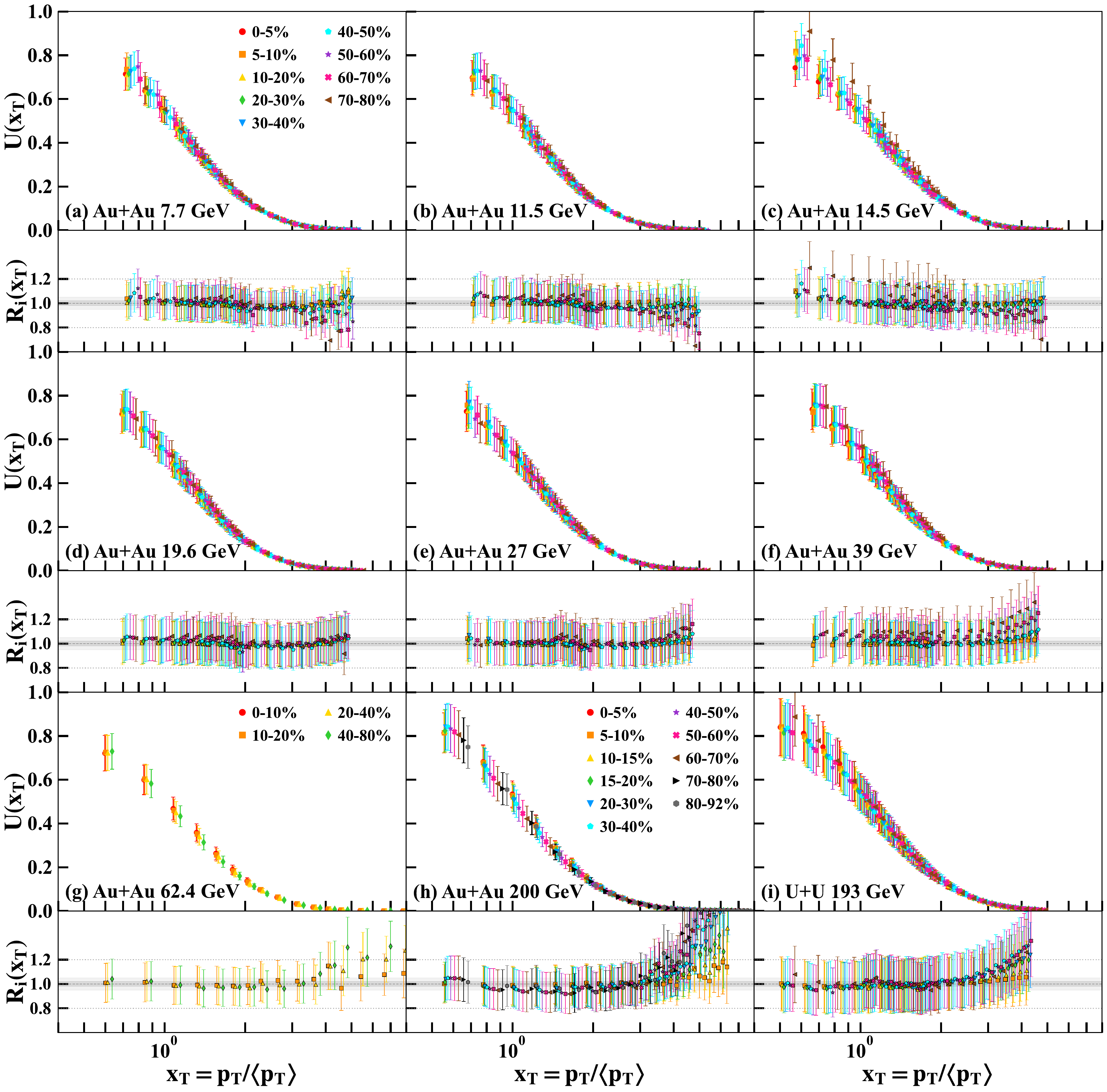}
\caption{For each row, the top panel shows the scaled spectra $U(x_{T})$ for $\pi^{+}$ in Au+Au collisions at 7.7~\cite{STAR:2017sal}, 11.5~\cite{STAR:2017sal}, 14.5~\cite{STAR:2019vcp}, 19.6~\cite{STAR:2017sal}, 27~\cite{STAR:2017sal}, 39~\cite{STAR:2017sal}, 62.4~\cite{STAR:2007zea}, and 200 ~\cite{PHENIX:2003iij} GeV, as well as in U+U collisions at 193 ~\cite{STAR:2022nvh} GeV, respectively. The bottom panel shows the corresponding ratios between $U_i(x_{T})$ from the $i$th centrality bin and that for the most central collisions in Au+Au collisions at 7.7–200 GeV and in U+U collisions at 193 GeV. The most central centrality is 0-5\% for all the collision systems except that it is 0-10\% in Au+Au collisions at 62.4 GeV. }
\label{figure1}
\end{figure*}

\begin{figure*}[!htbp]
\centering
\includegraphics[width=\textwidth]{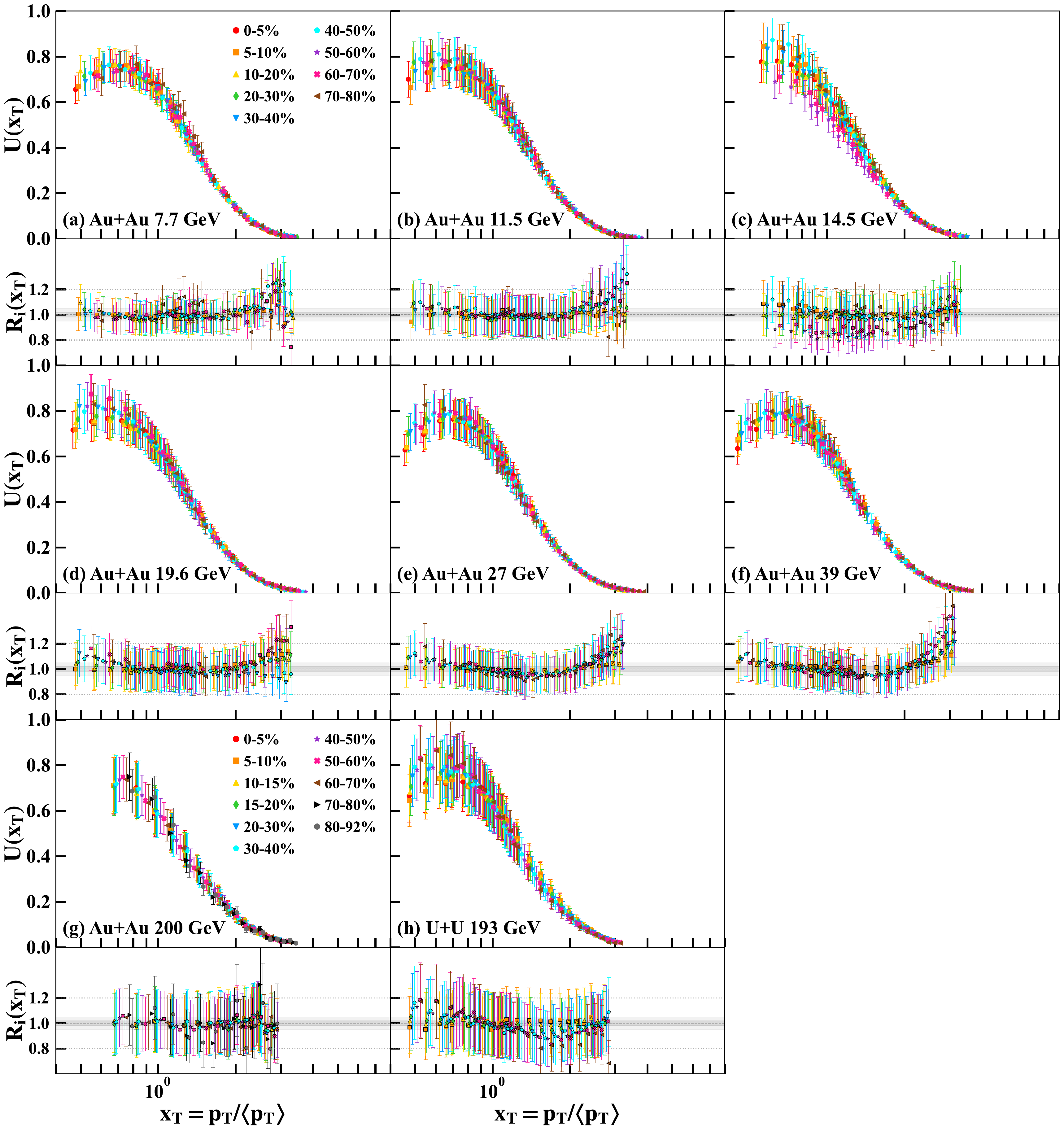}
\caption{For each row, the top panel shows the scaled spectra $U(x_{T})$ for $K^{+}$ in Au+Au collisions at 7.7~\cite{STAR:2017sal}, 11.5~\cite{STAR:2017sal}, 14.5~\cite{STAR:2019vcp}, 19.6~\cite{STAR:2017sal}, 27~\cite{STAR:2017sal}, 39~\cite{STAR:2017sal}, 62.4~\cite{STAR:2007zea}, and 200~\cite{PHENIX:2003iij} GeV, as well as in U+U collisions at 193~\cite{STAR:2022nvh} GeV. The bottom panel shows the corresponding ratios between $U_i(x_{T})$ from the $i$th centrality bin and that for the most central collisions in Au+Au collisions at 7.7–200 GeV and in U+U collisions at 193 GeV. The most central centrality is 0-5\% for all the collision systems.}
\label{figure2}
\end{figure*}

\begin{figure*}[!htbp]
\centering
\includegraphics[width=\textwidth]{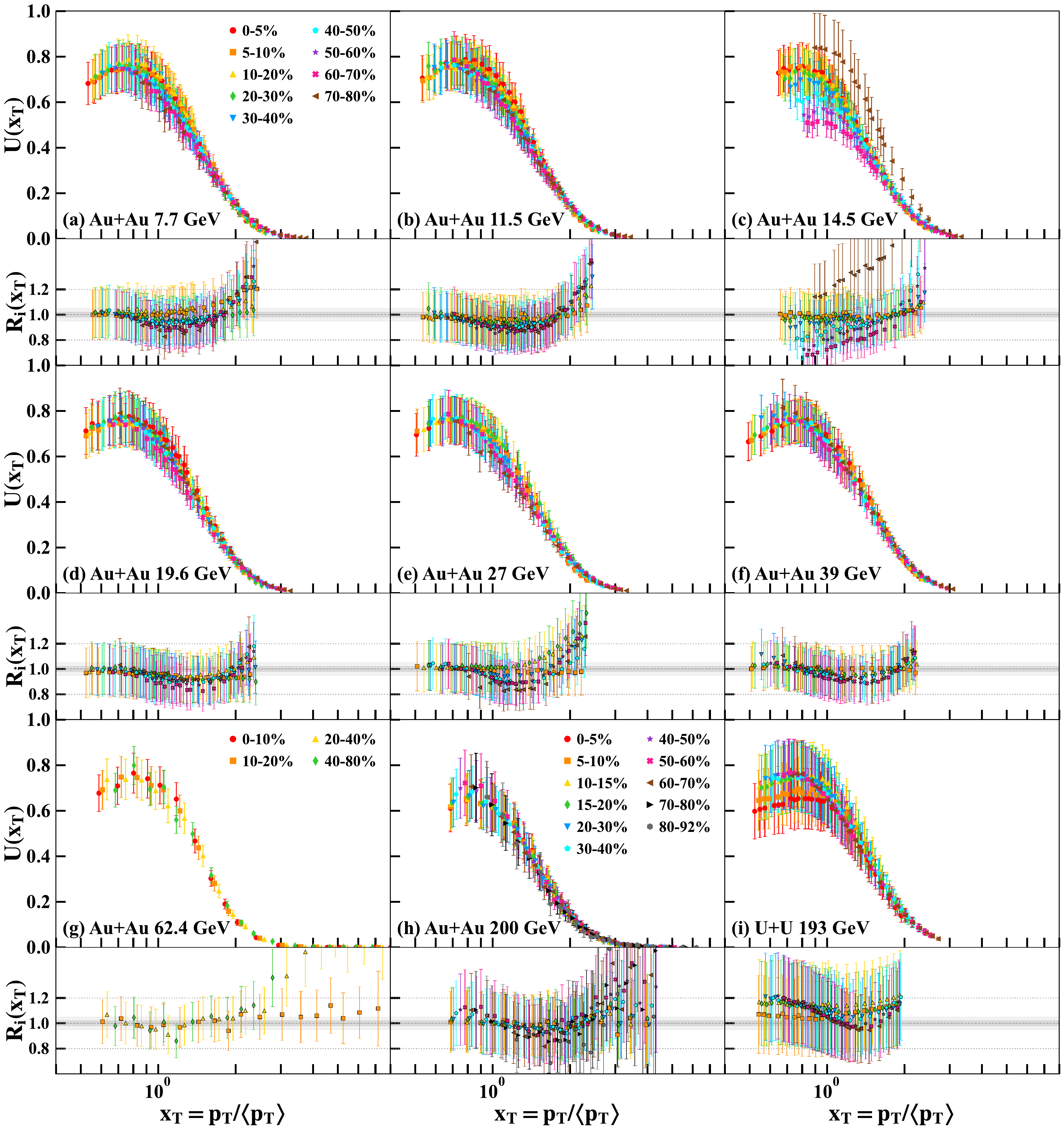}
\caption{For each row, the top panel shows the scaled spectra $U(x_{T})$ for $p$ in Au+Au collisions at 7.7~\cite{STAR:2017sal}, 11.5~\cite{STAR:2017sal}, 14.5~\cite{STAR:2019vcp}, 19.6~\cite{STAR:2017sal}, 27~\cite{STAR:2017sal}, 39~\cite{STAR:2017sal}, 62.4~\cite{STAR:2007zea}, and 200 ~\cite{PHENIX:2003iij} GeV, as well as U+U collisions at 193 ~\cite{STAR:2022nvh} GeV. The bottom panel shows the corresponding ratios between $U_i(x_{T})$ from the $i$th centrality bin and that for the most central collisions in Au+Au collisions at 7.7–200 GeV and in U+U collisions at 193 GeV. The most central centrality is 0-5\% for all the collision systems except that it is 0-10\% in Au+Au collisions at 62.4 GeV.}
\label{figure3}
\end{figure*}

The STAR and PHENIX collaborations at RHIC have published data for the transverse momentum spectra of pions, kaons, and protons at midrapidity across different centralities. We collect the data from Au+Au collisions at $\sqrt{s_{NN}}$ = 7.7~\cite{STAR:2017sal}, 11.5~\cite{STAR:2017sal}, 14.5~\cite{STAR:2019vcp}, 19.6~\cite{STAR:2017sal}, 27~\cite{STAR:2017sal}, 39~\cite{STAR:2017sal}, and 62.4~\cite{STAR:2007zea} GeV, 200~\cite{PHENIX:2003iij} GeV as well as from U+U collisions at $\sqrt{s_{NN}}$ = 193 ~\cite{STAR:2022nvh} GeV, and conduct the analyses. The scaled transverse momentum spectra $U(x_{\rm T})$, following the procedure in Sec. \ref{two}, for pions, kaons, and protons across collision centralities, systems and energies at RHIC are presented in Figs.~\ref{figure1},~\ref{figure2} and~\ref{figure3}, respectively. The results are shown in the same pattern. For a given collision energy, the scaled data points $U(x_{\rm T})$ corresponding to different centralities are distinguished by distinct colors and markers, shown in the top panel. Error bars represent the estimated uncertainties. To visualize the quality of the particle spectra scaling across the entire centrality range, the ratios 
$R_{i}(x_{\rm T})$, Eq.(\ref{ratioeq}), are also shown in the bottom panel. 

The scaled spectra $U(x_{\rm T})$ for pions across RHIC energies are shown in Fig.~\ref{figure1}. The scaled data points collapse onto a single curve at a given collision energy, revealing a particle transverse momentum spectra scaling independent of centrality and confirming the scaling findings at the LHC~\cite{ExTrEMe:2024fxt,Domingues:2025tkp}. Note that some deviations are observed in peripheral Au+Au collisions at $\sqrt{s_{NN}}$ = 14.5 GeV. The ratios $R_i(x_{\rm T})$ reveal more details at high $p_{\rm T}$ that the scaling quality is not as good as that at low $p_{\rm T}$. This behavior is attributed to the transition to a (semi)hard physics regime beyond hydrodynamics. Overall, a consistent pattern is observed across the RHIC energies. Similar results are found for kaons and protons, as shown in Figs.~\ref{figure2} and ~\ref{figure3}. 

A closer quantitative comparison among the particle species reveals subtle differences. The scaled spectra for pions adhere most precisely to the universal curve across all centralities, while kaons and protons show systematically larger deviations in peripheral collisions. This mass hierarchy, where heavier hadrons show relatively worse scaling, is consistent with phenomena observed at the LHC~\cite{Domingues:2025tkp} and can be understood within the collective expansion framework. In peripheral collisions, where radial flow is weaker, mass-dependent modifications of particle spectra become more apparent compared to central collisions. Consequently, heavier particles deviate more noticeably from the scaling in peripheral collisions~\cite{Bashir:2025hyt,Jiang:2013gxa}. 

\begin{figure}[!htbp]
\centering
\vspace{-0.5cm}

\begin{minipage}{1\linewidth}
\centering
\includegraphics[width=\linewidth]{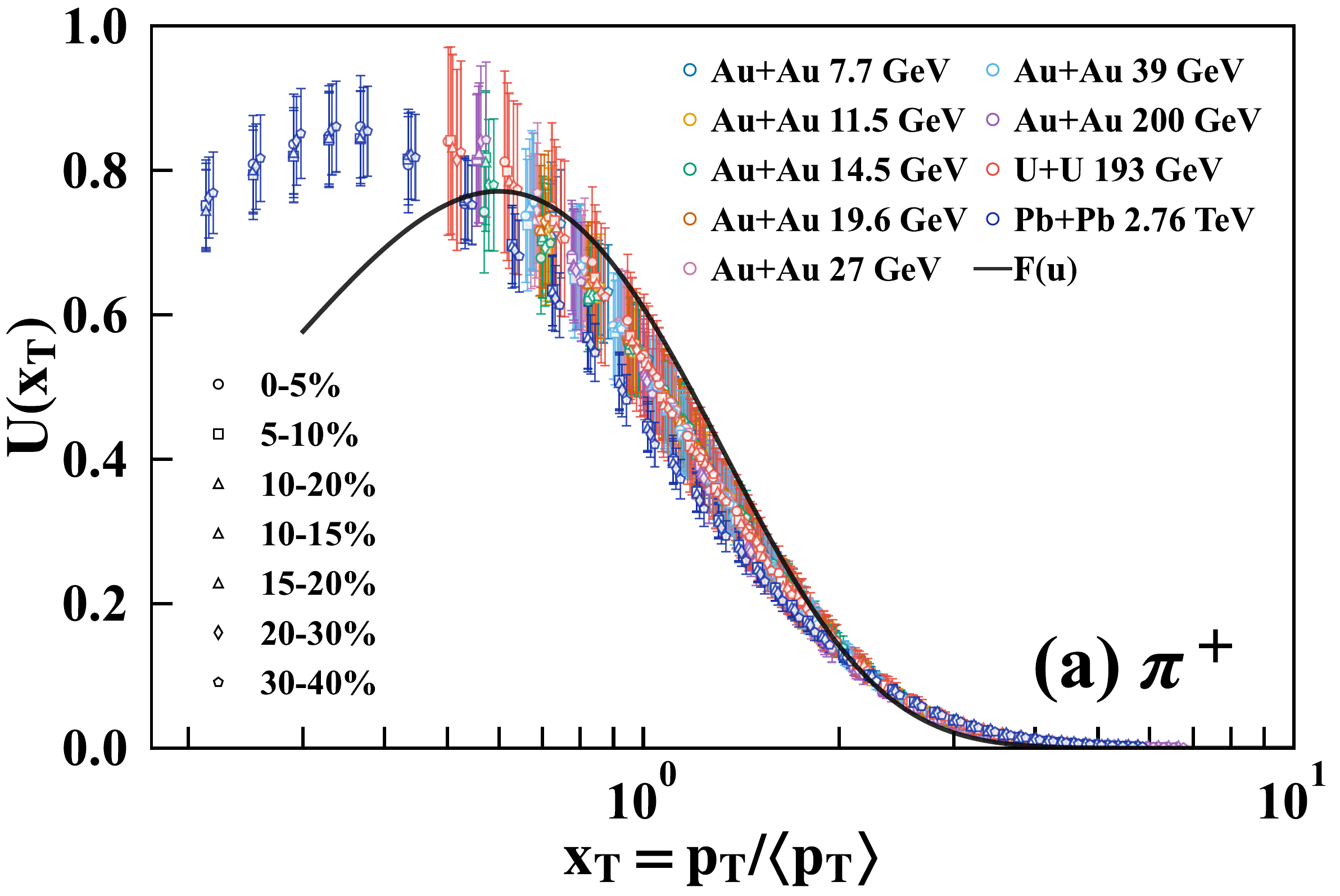}
\end{minipage}

\vspace{0.2cm} 

\begin{minipage}{1\linewidth}
\centering
\includegraphics[width=\linewidth]{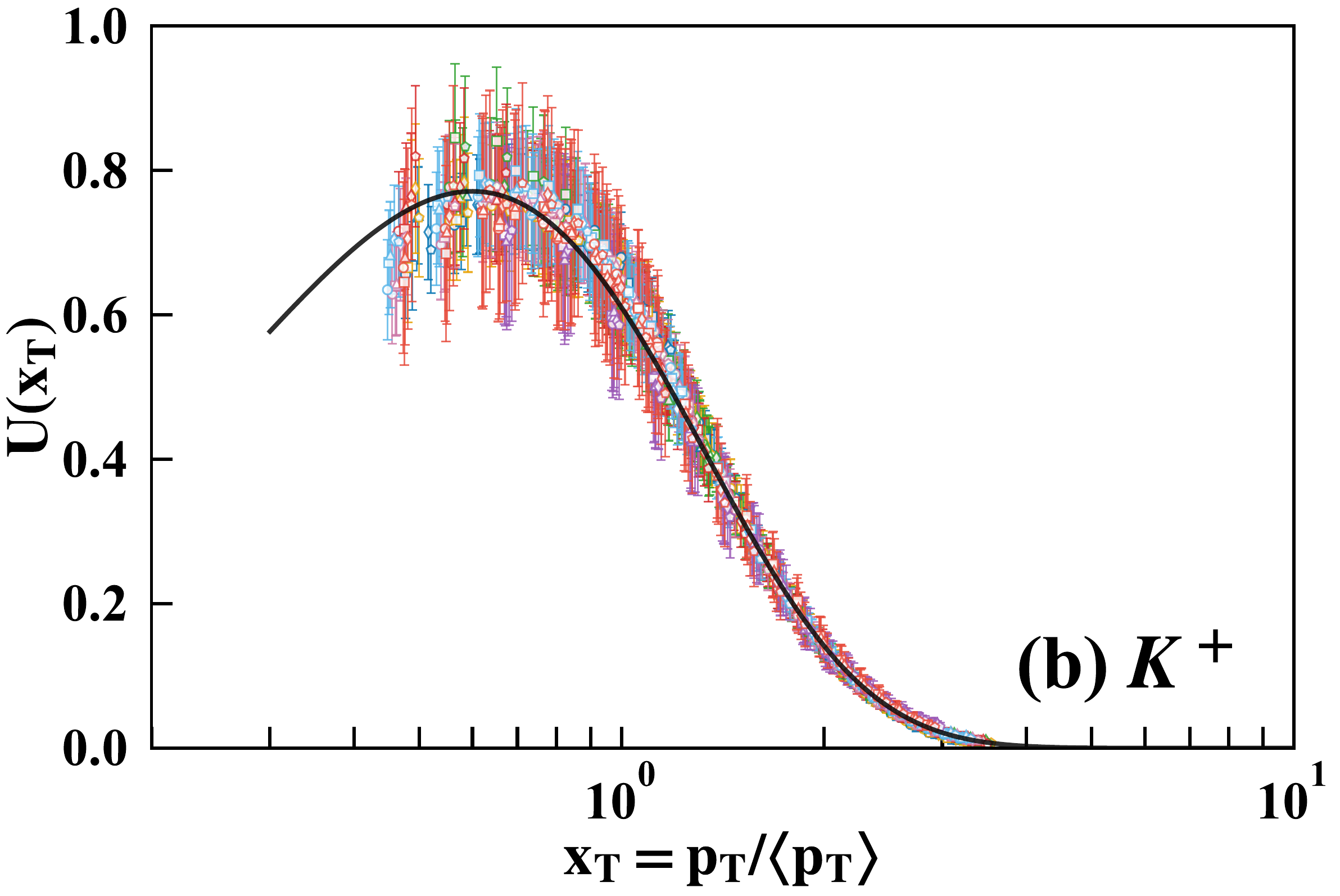}
\end{minipage}

\vspace{0.2cm}

\begin{minipage}{1\linewidth}
\centering
\includegraphics[width=\linewidth]{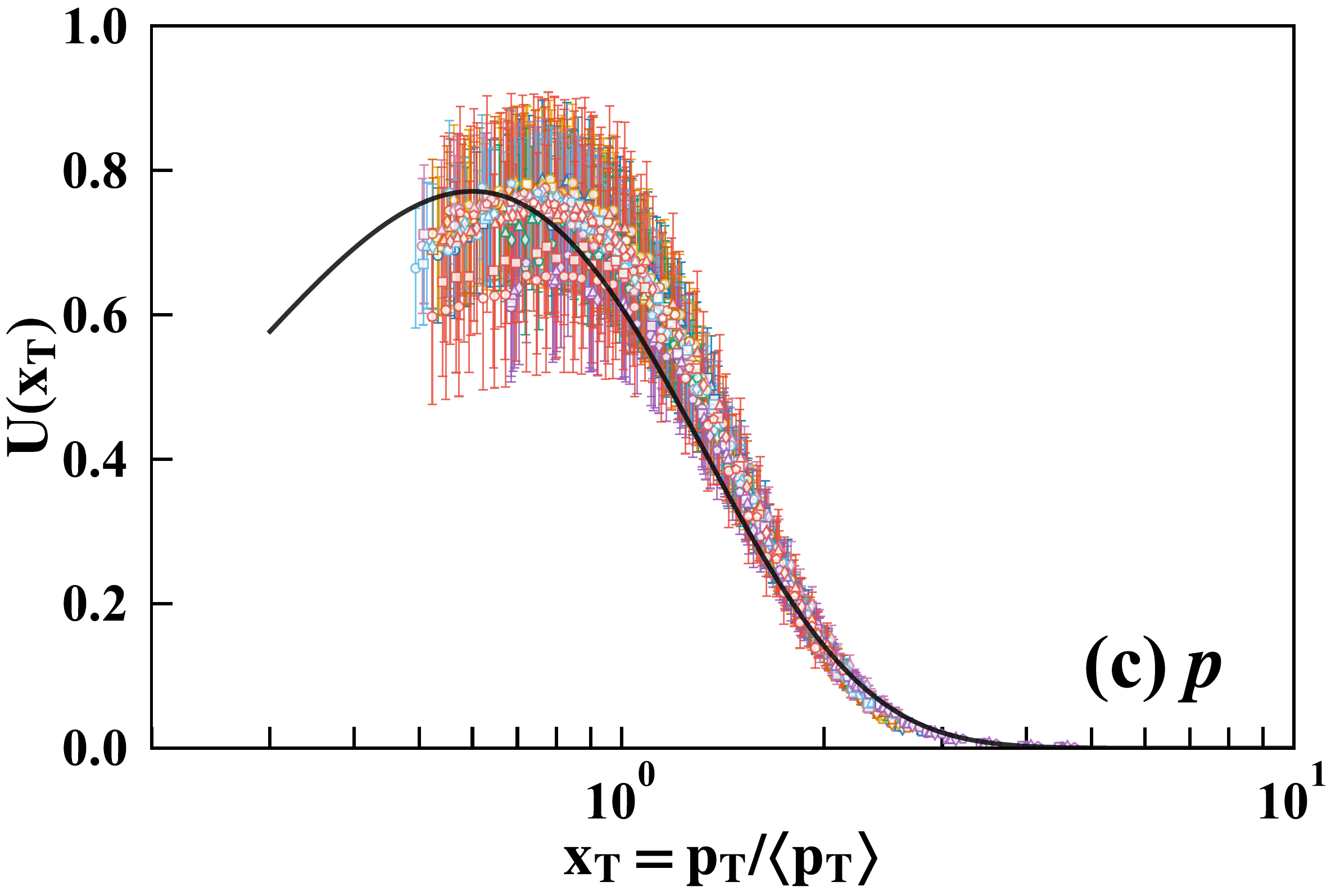}
\end{minipage}

\vspace{-0.2cm}
\caption{Panel (a) The scaled spectra $U(x_{\rm T})$ for $\pi^+$ in centrality interval 0--40\%, including the Au+Au collisions at $\sqrt{s_{NN}} = 7.7$--200 GeV, U+U collisions at 193 GeV, and Pb+Pb collisions at 2.76 TeV.
Panels (b, c) The scaled spectra $U(x_{\rm T})$ for $K^+$ and $p$ in Au+Au and U+U collisions at RHIC energies. The curves shown in panels (a, b, c) are the same and correspond to Eq. (\ref{scalingeq}).}
\label{figure4}

\end{figure}

We further investigate whether this scaling is universal and independent of the collision energy. Figure~\ref{figure4} shows the scaled spectra $U(x_{\rm T})$ for the same centrality interval (0–40\%) for different collision energies, including the results from LHC~\cite{ExTrEMe:2024fxt}. It is observed that the data points from 7.7 GeV to 200 GeV collapse onto a single curve, demonstrating the universality of this scaling. The persistence of scaling from 7.7 GeV to 200 GeV in Au+Au and U+U collisions, as well as the results from the LHC \cite{ExTrEMe:2024fxt}, establishes the scaled spectra as a baseline. Any clear deviation from the observed scaling could serve as a signal of different physics processes at the RHIC and LHC~\cite{ExTrEMe:2024fxt}.

\section{discussions}\label{four}

It has been shown that the particle spectra scaling exists and is independent of collision energies, collision centralities, and particle species. The ExTrEMe collaboration has demonstrated through event-by-event hybrid hydrodynamic simulations that this scaling phenomenon is naturally reproduced in fluid-dynamical models~\cite{ExTrEMe:2024fxt,Domingues:2025tkp,Domingues:2026mis,Domingues:2026xvs}. It was suggested that this scaling can serve as a probe of collective dynamics in heavy-ion collisions. However, a comprehensive study reveals that the hybrid hydrodynamic models have little flexibility in describing the scaled spectrum observables by varying the model parameters \cite{Domingues:2026mis}. This inspires us to pursue the origin of this universal scaling.

The hydrodynamic description provides a state-of-the-art framework for understanding collective expansion in heavy-ion collisions~\cite{Gale:2013da,Florkowski:2017olj,Zhao:2018lyf,STAR:2008ftz,Shi:2024pyz,Pang:2018zzo}. The produced matter is treated as a strongly coupled fluid that expands and cools in a hydrodynamic model. The dynamics of the system is governed by the conservation of energy and momentum, with transport coefficients such as shear and bulk viscosity accounting for dissipative effects. At late times, when the temperature becomes lower than the critical temperature, particles freeze-out via the Cooper-Frye prescription.

The Cooper-Frye formula is written as~\cite{Cooper:1974mv,Becattini:2013fla,Urmossy:2009jf,Che:2020fbz,Retiere:2003kf}
\begin{equation}
E\frac{d^3N}{dp^3} = \int_{\Sigma^\mu}d^3\sigma_\mu p^\mu f(x,p), \label{cooperfryeeq}
\end{equation}
where $\Sigma^\mu$ is the freeze-out hypersurface and $\sigma_\mu$ is its covariant normal vector. $f(x,p)$ is the Lorentz-invariant Boltzmann distribution at the kinetic freeze-out temperature $T_K$, given by~\cite{Florkowski:2004tn,Wong1994} 
\begin{equation}
f(x,p) = \frac{g\xi}{(2\pi)^3} \exp\left(-\frac{p^\mu u_\mu}{T_K}\right).
\end{equation}
Here, $g$ is the degeneracy factor and $\xi$ is the fugacity for the particle. The four-momentum $p^\mu$ takes the form
\begin{equation}
\begin{aligned}
p^\mu
&= (e,\boldsymbol{p})
= (e,\boldsymbol{p}_T,p_z) \\
&= \bigl(m_{\rm T}\cosh y,\; p_{\rm T}\cos\phi_p,\; p_{\rm T}\sin\phi_p,\; m_{\rm T}\sinh y\bigr),
\end{aligned}
\end{equation}
while the flow four-velocity $u^\mu(x)$ is
\begin{equation}
\begin{aligned}
u^\mu
&= \gamma_T(\cosh\eta,\boldsymbol{\beta},\sinh\eta) \\
&= \cosh\rho\bigl(\cosh\eta,\; \tanh\rho\cos\phi_b,\; \tanh\rho\sin\phi_b,\; \sinh\eta\bigr),
\end{aligned}
\end{equation}
with $\boldsymbol{\beta}$ the transverse flow velocity. The angles $\phi_p$ and $\phi_b$ correspond, respectively, to the azimuthal angles of the particle momentum and the transverse flow velocity relative to the $x$-axis.
$p^\mu u_\mu$ can be read as~\cite{Schnedermann:1993ws,Che:2020fbz}
\begin{equation}
\begin{aligned}
p^\mu u_\mu &= \gamma_T\left(e\cosh\eta-\boldsymbol{p}_T\cdot\boldsymbol{\beta}-p_z\sinh\eta\right), \\
            &= m_{\rm T}\cosh\rho\cosh(\eta-y)-p_{\rm T}\sinh\rho\cos(\phi_p-\phi_b).
\end{aligned}
\end{equation}
Under the assumption that $\Sigma^\mu$ is a constant-proper-time surface $\tau = \text{const}$, the normal element takes the explicit form
\begin{equation}
d^3\sigma_\mu = \left( \cosh\eta,\, 0,\, 0,\, -\sinh\eta \right) \tau r \, dr d\eta d\phi_b.
\end{equation}
Therefore we obtain
\begin{equation}
p^\mu d^3\sigma_\mu = \tau m_{\rm T}\cosh(\eta-y)d\eta r dr d\phi_b.
\end{equation}
Substituting the expressions for $p^\mu d^3\sigma_\mu$, $f(x,p)$ and $p^\mu u_\mu$ into the Cooper-Frye formula Eq. (\ref{cooperfryeeq}) gives
\begin{equation}
\begin{aligned}
E\frac{d^3N}{dp^3}
&= \frac{g\xi}{(2\pi)^3} \int \tau r dr d\eta d\phi_b \;
m_{\rm T}\cosh(\eta-y) \\
&\times \exp\left[-\frac{m_{\rm T}\cosh\rho\cosh(\eta-y)}{T_K}\right. \\
&\left.+ \frac{p_{\rm T}\sinh\rho}{T_K}\cos(\phi_p - \phi_b)\right].\label{cooperfryeeq1}
\end{aligned}
\end{equation}
The integral yields the modified Bessel function $I_0$~\cite{Schnedermann:1993ws,Che:2020fbz}
\begin{equation}
\int_{0}^{2\pi}\exp\left[\frac{p_{T}\sinh\rho}{T_{K}}\cos(\phi_{p}-\phi_{b})\right]d\phi_{b}=2\pi I_{0}\left(\frac{p_{T}\sinh\rho}{T_{K}}\right). \label{i0eq}
\end{equation}
Substituting Eq. (\ref{i0eq}) into Eq. (\ref{cooperfryeeq1}) and simplifying the prefactor, we obtain the invariant particle spectrum
\begin{equation}
\begin{aligned}
E\frac{d^3N}{dp^3}
&= \frac{g\xi}{(2\pi)^2}\int\tau r dr d\eta \, m_{\rm T}\cosh(\eta-y) \\
&\quad \times e^{-m_{\rm T}\cosh\rho\cosh(\eta-y)/T_K} I_0\left(\frac{p_{\rm T}\sinh\rho}{T_K}\right).
\end{aligned}
\end{equation}
Using the relation $E\frac{d^3N}{dp^3} = \frac{d^2N}{2\pi p_{\rm T} dp_{\rm T} dy}$, one has
\begin{equation}
\begin{aligned}
\frac{d^2N}{2\pi p_{\rm T}dp_{\rm T}dy}
&= \frac{g\xi}{(2\pi)^2}\int\tau r dr d\eta \, m_{\rm T}\cosh(\eta-y) \\
&\quad \times e^{-m_{\rm T}\cosh\rho\cosh(\eta-y)/T_K} I_0\left(\frac{p_{\rm T}\sinh\rho}{T_K}\right).
\end{aligned}
\end{equation}
At midrapidity ($y\approx 0$), we take the approximation that the main contribution is from $\eta\approx y$ and have $\cosh(\eta - y) = 1$. We also take the assumption that the transverse flow rapidity $\rho$ is independent of the radial coordinate $r$. Therefore, one has

\begin{equation}
\begin{aligned}
\left.\frac{d^{2}N}{2\pi p_{T}dp_{T}dy}\right|_{y=0}
&= \frac{g\xi}{8\pi^{2}}\tau\Delta\eta R^{2} m_{\rm T} \\
&\!\!\!\times e^{-m_{T}\cosh\rho/T_{K}}
I_0\left(\frac{p_{T}\sinh\rho}{T_{K}}\right).
\end{aligned}
\end{equation}
$\Delta \eta$ is a constant to take into count the contribution from $\eta$ space. Its value does not change our conclusion and will be shown in the following. In the midrapidity region, the rapidity dependence of the particle spectrum is very weak, thus the single differential particle spectrum is
\begin{equation}
\frac{dN}{dp_{\rm T}}=\frac{g\xi}{4\pi}\tau\Delta\eta\Delta yR^2p_{\rm T}m_{\rm T}e^{-m_{\rm T}\cosh\rho/T_K}I_0\left(\frac{p_{\rm T}\sinh\rho}{T_K}\right),
\end{equation}
where $\Delta y$ is the midrapidity width.

To obtain analytic formula for deciphering the origin of particle transverse momentum spectra scaling, we take the massless particle limit and have $m_{T}\approx p_{T}$. Therefore
\begin{equation}
\frac{dN}{dp_{T}}\approx\frac{g\xi}{4\pi}\tau\Delta\eta\Delta yR^{2}p_{T}^{2}e^{-p_{T}\cosh\rho/T_{K}}I_{0}\left(\frac{p_{T}\sinh\rho}{T_{K}}\right).
\end{equation}

Using the asymptotic approximation $I_0(z) \approx e^z/\sqrt{2\pi z}$ for $z \equiv p_{T}\sinh\rho/T_{K}$, which is justified for the majority of experimental data points, we obtain
\begin{equation}
\begin{aligned}
\frac{dN}{dp_{\rm T}}
&\approx \frac{g\xi}{4\pi}\tau\Delta\eta\Delta yR^2
         \frac{p_{\rm T}^{3/2}}{\sqrt{2\pi\sinh\rho/T_K}} \\
&\quad \times \exp\left[-\frac{p_{\rm T}(\cosh\rho-\sinh\rho)}{T_K}\right].
\end{aligned}
\end{equation}
Defining an effective temperature
\begin{equation}
T_{\mathrm{eff}} \equiv \frac{T_K}{\cosh \rho - \sinh \rho} = T_K e^{\rho},
\end{equation}
the expression of \(\frac{dN}{dp_{\rm T}}\) simplifies to
\begin{equation}
\frac{dN}{dp_{\rm T}} = C_0 \, p_{\rm T}^{3/2} \, e^{-p_{\rm T}/T_{\mathrm{eff}}}, \label{spectrumeq}
\end{equation}
where $C_0 = \frac{g\xi}{4\pi} \tau \Delta \eta \Delta y R^2 \, \frac{1}{\sqrt{2\pi \sinh \rho / T_K}}$ is a constant.

Now we can calculate the mean transverse momentum $\langle p_{\rm T}\rangle$ and average total  multiplicity $N$ for a given particle. The mean transverse momentum is 
\begin{equation}
\begin{aligned}
\langle p_{\rm T}\rangle 
&= \frac{\int_0^\infty p_{\rm T} \frac{dN}{dp_{\rm T}} dp_{\rm T}}{\int_0^\infty \frac{dN}{dp_{\rm T}} dp_{\rm T}}
= \frac{\int_0^\infty p_{\rm T}^{5/2} e^{-p_{\rm T}/T_{\rm eff}} dp_{\rm T}}{\int_0^\infty p_{\rm T}^{3/2} e^{-p_{\rm T}/T_{\rm eff}} dp_{\rm T}} \\
&= \frac{T_\mathrm{eff}^{7/2}\Gamma(7/2)}{T_\mathrm{eff}^{5/2}\Gamma(5/2)}
= \frac{5}{2}T_{\mathrm{eff}}.
\end{aligned}
\end{equation}
Therefore, the relation between $\langle p_{\rm T}\rangle$ and effective temperature \(T_{\mathrm{eff}}\) is
\begin{equation}
T_{\mathrm{eff}} = \frac{2}{5}\langle p_{\rm T}\rangle. \label{teff}
\end{equation}

Similarly, the average total multiplicity is
\begin{equation}
N = \int_0^\infty \frac{dN}{dp_{\rm T}} dp_{\rm T}
= C_0 \int_0^\infty p_{\rm T}^{3/2} e^{-p_{\rm T}/T_{\rm eff}} dp_{\rm T}.
\end{equation}
Using the relation Eq. (\ref{teff}) and introducing the scaled variable $u = p_{\rm T}/\langle p_{\rm T}\rangle$, we obtain
\begin{equation}
N = C_0 \langle p_{\rm T}\rangle^{5/2} \int_0^\infty u^{3/2} e^{-5u/2} du.
\end{equation}
\(C_0\) can be expressed as
\begin{equation}
C_0 = \frac{N}{\langle p_{\rm T}\rangle^{5/2}} \cdot \frac{25}{3} \sqrt{\frac{5}{2\pi}}.
\end{equation}
Substituting \(C_0\) into Eq. (\ref{spectrumeq}), one obtains
\begin{equation}
\frac{dN}{dp_{\rm T}} = \frac{N}{\langle p_{\rm T}\rangle} \frac{25}{3} \sqrt{\frac{5}{2\pi}} \, u^{3/2} e^{-5u/2}.
\end{equation}
Thus the scaled particle transverse momentum spectrum is
\begin{equation}
U(x_{\rm T})=\frac{\langle p_{\rm T}\rangle}{N}\frac{dN}{dp_{\rm T}} = F(u),
\end{equation}
where the scaling function \(F(u)\) is
\begin{equation}
F(u) = \frac{25}{3} \sqrt{\frac{5}{2\pi}} \, u^{3/2} e^{-5u/2}. \label{scalingeq}
\end{equation}

Therefore, the universal scaling behavior observed from particle transverse momentum spectra in heavy-ion collisions can be naturally understood with the Cooper-Frye formula using the assumptions made in the derivation. Equation (\ref{scalingeq}) does not depend on the collision centralities, collision energies, and collision systems as long as the Cooper-Frye formula holds. It is not surprising to see that the particle transverse momentum spectra scaling has the particle mass dependence from data \cite{ExTrEMe:2024fxt,Domingues:2025tkp,Domingues:2026mis,Domingues:2026xvs} since we take the massless particle limit. 
In Fig. \ref{figure4}, the curves from Eq. (\ref{scalingeq}) are also shown, which can describe the general features of the particle transverse momentum spectra scaling from experimental data. Equation (\ref{scalingeq}) may also explain why the hybrid hydrodynamic models have little flexibility in describing the scaled spectrum observables by varying model parameters. Because the universal scaling is strongly relevant to the Cooper-Frye formula which is used for the particle freeze-out in hydrodynamic model. Our conjecture can be investigated in the hybrid hydrodynamic model but it is out of the scope of the present paper.

As mentioned before, there is another particle transverse momentum spectra scaling discovered earlier, i.e., Hwa-Yang scaling.  They introduced a scaling variable $z=p_{\rm T}/{K}$, where $K$ is a scaling parameter that depends on both the collision energy and centrality. The scaled spectrum is defined as
\begin{equation}\label{eq2}
\Phi(z)=\left.A \cdot \frac{1}{2 \pi p_{\rm T}} \frac{d^{2} N}{d p_{\rm T} d y}\right|_{z=\frac{p_{\rm T}}{K}},
\end{equation}
where $A$ is a normalization factor. By adjusting values of $K$ and $A$ for each particle transverse momentum spectrum, the scaled particle spectra at different energies and centralities all collapse onto the same universal curve $\Phi(z)$. To eliminate the arbitrariness in choosing the reference spectrum, Hwa and Yang further introduced a KNO-like scaling variable $u=z/\left \langle z \right \rangle = p_{\rm T}/\left \langle p_{\rm T} \right \rangle$, and the corresponding normalized scaling function is defined as
\begin{equation}
\Psi(u) = \frac{\langle z \rangle^2 \Phi(\langle z \rangle u)}{\int \Phi(z) z dz}.\label{yangeq0}
\end{equation}
This function $\Psi(u)$ is independent of both collision energy and centrality, thus providing a universal scaling for the particle spectra from different collision centralities and systems~\cite{Hwa:2002tu,Hwa:2003bn,Hwa:2003zp,Zhu:2006gv,Zheng:2008zzg,Zhu:2008zzd,Zhu:2008uu,Zhang:2007st,Zhang:2014dna,Zhang:2015wea,Yang:2017cup,Wang:2018pbh,Liu:2019fdp}. 

Now we turn to deriving the relation between the scaling found by ExTrEMe collaboration and the Hwa-Yang scaling. The invariant yield distribution can be rewritten from Eq. (\ref{eq2}) as
\begin{equation}
\frac{d^{2} N}{d p_{\rm T} d y}=\frac{2 \pi p_{\rm T}}{A} \Phi(z), \quad z=\frac{p_{\rm T}}{K}. 
\end{equation}
At midrapidity, the rapidity dependence of the particle transverse momentum spectrum is very weak, so one has
\begin{equation}
\frac{d N}{d p_{\rm T}} \approx \Delta y \cdot \frac{ 2 \pi p_{\rm T}}{A} \Phi(z). \label{yangeq1}
\end{equation}
The average total particle multiplicity $N$ for a given particle is then calculated by
\begin{equation}
N=\int_{0}^{\infty} \frac{d N}{d p_{\rm T}} d p_{\rm T}=\int_{0}^{\infty} \Delta y \cdot \frac{2 \pi p_{\rm T}}{A} \Phi(z) d p_{\rm T}.
\end{equation}
Since $p_{\rm T}=Kz$, then $dp_{\rm T}=Kdz$, one has
\begin{equation}
N=\Delta y\cdot\frac{2\pi}{A}K^2\int_0^\infty\Phi(z)zdz. \label{yangeq2}
\end{equation}
It is obvious that
\begin{equation}
\langle p_{\rm T}\rangle = K \langle z \rangle.\label{yangeq3}
\end{equation}

Substituting Eqs. (\ref{yangeq1}, \ref{yangeq2}, \ref{yangeq3}) into the scaled particle transverse momentum spectrum $U(x_{\rm T})$ i.e., Eq. (\ref{eq1}), and using the relations $p_{\rm T} = \langle p_{\rm T}\rangle x_{\rm T} = K\langle z\rangle x_{\rm T}$, $z = \langle z\rangle x_{\rm T}$, we obtain
\begin{align}
U(x_{\rm T}) &= \frac{K\langle z\rangle}{K^2\int\Phi(z)zdz}\cdot(K\langle z\rangle x_{\rm T})\Phi(\langle z\rangle x_{\rm T}) \notag \\
&= \frac{x_{\rm T}\langle z\rangle^2\Phi(\langle z\rangle x_{\rm T})}{\int\Phi(z)zdz}.
\end{align}

Noting the normalized Hwa-Yang scaling function Eq. (\ref{yangeq0}) and $u=p_{\rm T}/\left \langle p_{\rm T} \right \rangle = x_{\rm T}$, we obtain a simple relation
\begin{equation}\label{eq3}
U(x_{\rm T}) = x_{\rm T} \Psi(x_{\rm T}).
\end{equation}
It demonstrates that the two scaling functions are mathematically equivalent, which is another significant contribution of this work. It is worth noting that the particle transverse momentum spectra scaling found by the ExTrEMe collaboration has more intuitive physical meaning.

\begin{figure}[htbp]
    \centering
    \begin{minipage}[b]{0.46\textwidth}
        \centering
        \includegraphics[width=\textwidth]{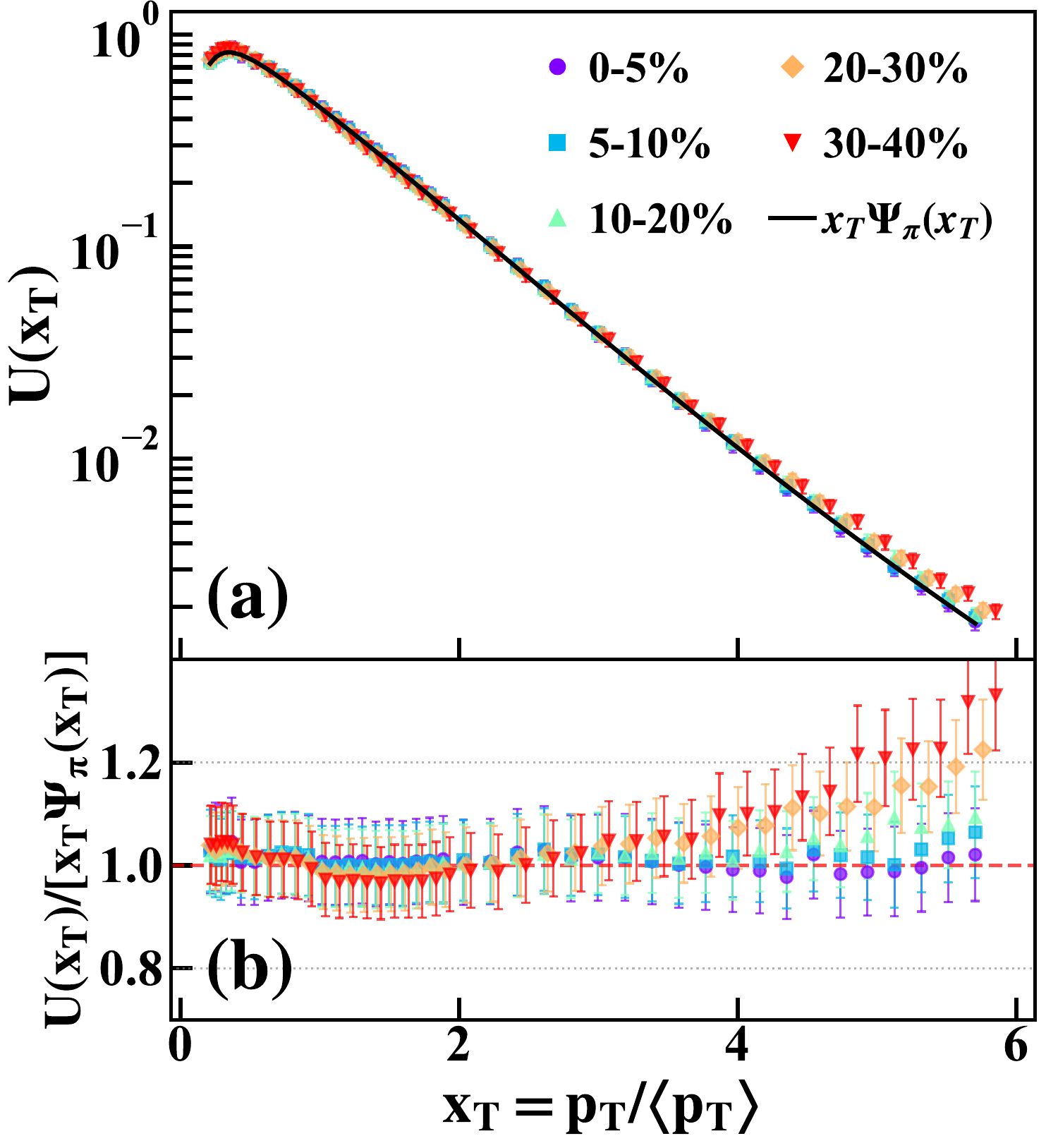}
    \end{minipage}
    \caption{Panel (a) shows a direct comparison between $U(x_{\rm T})$ (symbols) and $x_{\rm T}\Psi(x_{\rm T})$ (solid curve) for various centrality classes using pion spectra from Pb+Pb collisions at $\sqrt{s_{NN}} = 2.76$ TeV~\cite{ALICE:2013hur}. Panel (b) presents the ratio $R=U(x_{\rm T})/[x_{\rm T}\Psi_\pi(x_{\rm T})]$ for five different centrality bins. The scaling function $\Psi_\pi(x_{\rm T})$ is taken from Ref. \cite{Wang:2018pbh}.}
    \label{figure5}
\end{figure}

Figure~\ref{figure5} presents a direct comparison of both sides of Eq. (\ref{eq3}) using the pion data from Pb+Pb collisions at $\sqrt{s_{NN}} = 2.76$ TeV~\cite{ALICE:2013hur} and the scaling function $\Psi_\pi(x_{\rm T})$ obtained in Ref.~\cite{Wang:2018pbh}. Panel (a) displays $U(x_{\rm T})$ (symbols) for various centrality classes together with $x_{\rm T}\Psi_\pi(x_{\rm T})$ (curve), while panel (b) shows their ratios. The ratios remain close to unity across all centralities, confirming the relation Eq. (\ref{eq3}).

\section{conclusion}\label{five}
In this paper, we have systematically investigated the scaling properties of the transverse momentum spectra for pions, kaons, and protons in Au+Au collisions from $\sqrt{s_{NN}} = 7.7$ GeV to 200 GeV and in U+U collisions at 193 GeV. Our results demonstrate that the scaled spectra $U(x_{\rm T})$ exhibit a centrality-independent shape, confirming that the universality previously observed at the LHC by the ExTrEMe collaboration extends to the lower energy regime of RHIC. However, a closer examination of the data uncovers characteristic deviations from this ideal scaling. We observe that the scaling breaks down in the high-$p_{\rm T}$ region and in peripheral collisions, signaling a transition from the hydrodynamic flow regime (soft) to a (semi)hard physics regime. Furthermore, a distinct mass hierarchy is observed. In contrast to pions, which adhere most precisely to the universal curve, kaons and protons show systematically larger deviations in peripheral collisions. This mass-dependent modification is consistent with a reduced radial flow strength in non-central collisions.

Theoretically, we provide a natural explanation for these phenomena using the Cooper-Frye formula with certain assumptions. A universal scaling function is derived. We show that the scaling is strongly relevant to the particle freeze-out mechanism by using our results as well as the results from a comprehensive study using the hybrid hydrodynamic models conducted by the ExTrEMe collaboration \cite{Domingues:2026mis}. Additionally, we establish the relation between the scaling found by the ExTrEMe collaboration and the Hwa-Yang scaling, demonstrating that the two scalings are mathematically equivalent. It is interesting to decipher the origin of particle transverse momentum spectra scaling in heavy-ion collisions even though the Cooper-Frye formula may not be the unique explanation.

\bibliography{reference.bib}

\end{document}